\documentclass[journal]{IEEEtran}
\usepackage{latexsym}
\usepackage{amssymb}
\usepackage{amsmath} 
\usepackage{booktabs}
\usepackage{enumerate}
\usepackage{graphicx}
\usepackage{subfigure}
\usepackage{xspace}
\usepackage{float}
\usepackage{bbm}
\usepackage{bm}
\usepackage{multirow}
\usepackage{color}
\usepackage{framed}
\usepackage{stfloats}
\usepackage{iitem}
\usepackage{pifont}
\usepackage{multibib}

\usepackage{array}
\usepackage{longtable}
\usepackage{colortbl}
\usepackage{colortab}
\usepackage{arydshln}
\usepackage{CJKutf8}

\usepackage{ragged2e}
\usepackage{diagbox}
\usepackage{algorithm}
\usepackage{algpseudocode}

\usepackage{pifont}

\ifCLASSINFOpdf
\else
\fi
\usepackage{hyperref}

\begin{document}
%
\title{Can Linguistic Knowledge Improve Multimodal Alignment in Vision-Language Pretraining?}
%
%
%

\author{Fei~Wang,
    Liang~Ding,~\IEEEmembership{Member,~IEEE},
    Jun~Rao,
    Ye~Liu,
    Li~Shen,
    Changxing~Ding,~\IEEEmembership{Member,~IEEE}
\thanks{
Fei~Wang, Ye~Liu, and Changxing~Ding are with the School of Future Technology, South China University of Technology, Guangzhou 511436, China~(e-mail:~\href{mailto: ft\_feiw@mail.scut.edu.cn}{ft\_feiw@mail.scut.edu.cn}; \href{mailto: yliu03@scut.edu.cn}{yliu03@scut.edu.cn}; \href{mailto: chxding@scut.edu.cn}{chxding@scut.edu.cn}).
}
\thanks{
Liang~Ding is with the School of Computer Science, University of Sydney, NSW 2006, Australia (e-mail: \href{mailto: liangding.liam@gmail.com}{liangding.liam@gmail.com})
}
\thanks{
Jun~Rao is with the School of Computer Science and Technology, Harbin Institute of Technology, Shenzhen 518055, China (e-mail:~\href{mailto: rao7jun@gmail.com}{rao7jun@gmail.com}).
}
\thanks{Li~Shen is with the JD Explore Academy at JD.com, Beijing 100176, China~(e-mail:~\href{mailto: mathshenli@gmail.com}{mathshenli@gmail.com}).}
}

%
%

\markboth{Journal of \LaTeX\ Class Files,~Vol.~14, No.~8, August~2015}%
{Shell \MakeLowercase{\textit{et al.}}: Bare Demo of IEEEtran.cls for IEEE Journals}
%




\maketitle

\begin{abstract}
The multimedia community has shown a significant interest in perceiving and representing the physical world with multimodal pretrained neural network models, and among them, the visual-language pertaining (VLP) is, currently, the most captivating topic. 
The common practice for pretraining the visual-language backbone is supervising the training process with paired image-text data.
However, there have been few endeavors dedicated to the exploration of 1) whether essential linguistic knowledge (e.g., semantics and syntax) can be extracted during VLP, and 2) how such linguistic knowledge impact or enhance the multimodal alignment.
In response, here we aim to elucidate the impact of comprehensive linguistic knowledge, including semantic expression and syntactic structure, on multimodal alignment.
Specifically, we design and release the SNARE, the first large-scale multimodal alignment probing benchmark, to detect the vital linguistic components, e.g., lexical, semantic, and syntax knowledge, containing four tasks: Semantic structure, Negation logic, Attribute ownership, and Relationship composition. 
Based on our proposed probing benchmark, our holistic analyses of five advanced VLP models (i.e., BLIP, CLIP, Flava, X-VLM, and BLIP2) illustrate that the VLP model: 
\textit{i)} shows insensitivity towards complex syntax structures and relies on content words for sentence comprehension;
\textit{ii)} demonstrates limited comprehension of combinations between sentences and negations;
\textit{iii)} faces challenges in determining the presence of actions or spatial relationships within visual information and struggles with verifying the correctness of triple combinations. 
Given the above findings, we suggest that, to improve the multimodal alignment, 1) using the large generative language model as the language backbone in VLP to understand complex sentences; 2) establishing high-quality datasets by highlighting the content words and using simple syntax (e.g., short-distance semantic composition) to improve multimodal alignment; and 3) incorporating more fine-grained visual knowledge (e.g., spatial relationships) into pretraining objectives. 
We make our benchmark and code available at \url{https://github.com/WangFei-2019/SNARE/}.
\end{abstract}


%
\IEEEpeerreviewmaketitle

\begin{figure*}[t!]
\centering
\includegraphics[scale=0.52]{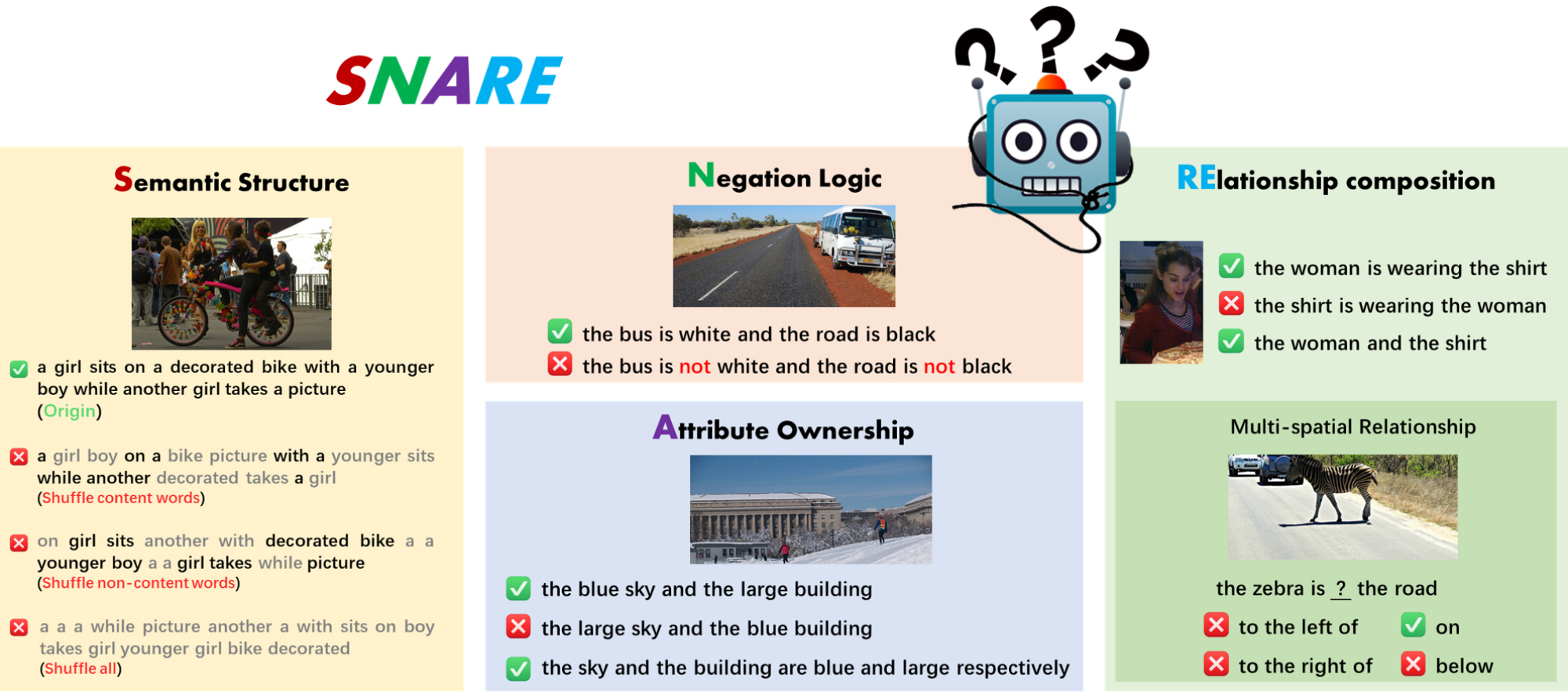}
\caption{\textbf{Samples in the SNARE benchmark for each task}.
\includegraphics[height=1.2em]{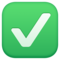} and \includegraphics[height=1.2em]{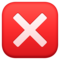} indicate the matching or not between the image and text. 
In the  Multi-spatial Relationship, a sub-task of Relationship Composition, the options need to fill in `\underline{ ? }' to form a complete sentence and try to align with the image.}
\label{fig:overview}
\end{figure*}

\section{Introduction}
\IEEEPARstart{V}{isual-language} pretraining (VLP) is a technology that aims to learn and align multimodal knowledge from large-scale pretraining datasets using carefully designed architecture~\cite{visualbert, vilbert, radford2021learning, chen2022pali}. 
After fine-tuning the pretrained VLP models, they exhibit better comprehension, cognitive, and reasoning ability in downstream tasks, such as multimodal machine translation~\cite{caglayan2019probing,li2022vision}, image-text retrieval~\cite{rao2022does, rao2023DCD}, multimodal reasoning~\cite{bugliarello2021multimodal, iki2021effect}. 
The multimodal alignment knowledge from pretraining image-text pairs is the key factor determining VLP models' generalization and downstream performance.
Recently, the development of large language models (LLMs)~\cite{chatgpt, touvron2023llama} have propelled VLP methods to a new paradigm, commonly known as multimodal large language models~(MLLMs)~\cite{yin2023survey, li2022blip}. 
However, despite being built upon LLMs that contain rich linguistic knowledge, MLLMs still face limitations in recognizing complex visual content and generating logically coherent responses conditioned on the vision information~\cite{yin2023lamm}. 
This has sparked our curiosity about the potential \textbf{influence of linguistic knowledge on multimodal alignment in VLP}, as it plays a crucial role in improving visual understanding, cross-modality reasoning, and generation.

Compositionality is a fundamental presupposition to robustly and accurately represent, understand, reason, and generate linguistic knowledge~\cite{cresswell2016logics}, where syntax governs the rule of compositionality and semantics response the outcomes~\cite{hsieh2023sugarcrepe}.
Prior studies of VLP probing~\cite{choi2012context, cao2020behind} primarily concentrate on the representation and richness of semantic knowledge within the models and can be divided into three primary categories.
First, probing whether knowledge from multimodal training is better than unimodal one:
Yun~\textit{et al.}~\cite{yun2021does} studied whether VLP improves linguistic knowledge comprehension. Salin~\textit{et al.}~\cite{salin2022vision} compared knowledge learned from visual/ textual models to multimodal models. 
Second, probing whether VLP models can infer semantic relationships in images and text:
Shekhar~\textit{et al.}~\cite{shekhar2017foil} and Hendricks~\textit{et al.}~\cite{hendricks2021probing} studied noun and verb comprehension, respectively. R{\"o}sch~\textit{et al.}~\cite{rosch2022probing} focused on learning and reasoning about location information in VLP models. Thrush~\textit{et al.}~\cite{thrush2022winoground} formed a manual dataset named Winoground to probe VLP models' ability on recognizing similar semantics.
The third category involves exploring the complementarity of different modality knowledge. 
Liu~\textit{et al.}~\cite{liu2022things} and Alper~\textit{et al.}~\cite{alper2023bert} studied the complementarity of visual knowledge to textual knowledge.
These methods primarily rely on the model's comprehension of semantic features in vision or language, and limited research has been dedicated to exploring the influence of diverse linguistic knowledge, particularly syntax, on multimodal alignment.

To examine the influence of semantics and syntax on multimodal alignment, we design and introduce \textsc{SNARE}, a pioneer multimodal alignment probing benchmark, which encompasses four tasks: a) \textbf{Semantic Structure}, b) \textbf{Negation Logic}, c) \textbf{Attribute Ownership}, and d) \textbf{Relationship Composition}. 
In addition to exploring rich syntactic knowledge, SNARE provides a comprehensive investigation at the semantic level compared to previous studies and we show the comparison with previous approaches in Tab.~\ref{tab:methods}. For a better understanding of our SNARE benchmark, we show samples of tasks in Fig.~\ref{fig:overview}.

In a) \textbf{Semantic Structure}, we partition words based on the part of speech and then separately shuffle the positions of content words (words with specific meanings) and others to disrupt compositionality, including semantics and syntax. It aims to investigate the VLP model's dependence on a particular type of words and whether it exhibits sensitivity to order.
b) \textbf{Negation Logic} adds negation words (``\textit{not}'') to sentences to test the model's understanding of negation logic.
In c) \textbf{Attribute Ownership}, we introduce sentences with different syntactic expressions, including short- and long-distance semantic combinations, e.g., short one (distance between ``white/ black'' and ``cat/ dog'' is 1): ``\textit{the white cat and the black dog}'' vs. long one (distances of ``white/ black'' and ``cat/ dog'' become 5/ 4, respectively): ``\textit{the cat and the dog are white and black, respectively}''. Also, we count sentences that contain the exact words but have different semantic combinations that do not match the image, such as ``\textit{the black dog and the white cat}''. This Attribute Ownership aims to evaluate the model's understanding of different syntactic forms of expression.
In d) \textbf{Relationship Composition}, we construct sentences using triplet that include two objects and a relationship. Specifically, we generate sentences with correct syntactic expressions (e.g., ``\textit{the girl is wearing the shirt}''), sentences with incorrect expressions where the positions of the two objects are exchanged (e.g., ``\textit{the shirt is wearing the girl}''), and sentences without the relationship word (e.g., ``\textit{the girl and the shirt}''). These sentences are used to investigate whether the VLP model comprehends the relationship element in sentences accurately and whether it understands the syntactic combination of object elements and relationship elements within the sentences.
In Relationship Composition, we further extract information to construct a sub-test set called ``Multi-Spatial Relationship'' to explore the model's understanding of spatial relationships between objects.

With the carefully designed probing benchmark SNARE, we evaluate four state-of-the-art VLP models, including BLIP~\cite{li2022blip}, CLIP~\cite{clip}, Flava~\cite{singh2022flava} and X-VLM~\cite{zeng2021multi}. We also extend SNARE to be compatible with MLLMs and test BLIP2~\cite{li2023blip2}. 
Through extensive analyses and experiments, we conclude some consistent and important findings: 
\begin{itemize}
    \item[\ding{182}] On the lexical level, VLP models prefer simple content words rather than more precise and complete function words that could make the sentence semantically legal (e.g., ``\textit{girl wearing shirt}'' instead of ``\textit{the girl is wearing the shirt}'').
    \item[\ding{183}] On the syntactic level, VLP models can easily comprehend short-distance syntactic combinations and simple relations (e.g., ``\textit{the white cat and the black dog}''), while they have difficulty in understanding long and relatively complicated syntactic combinations (e.g.,
    ``\textit{The cat and the dog are white and black, respectively}'').
    \item[\ding{184}] On the semantic level, VLP models \textbf{encounter difficulties in} a) comprehending the semantics of negation (e.g., ``\textit{is not}''), b) precisely discerning spatial relations between objects (particularly ``\textit{left}'' and ``\textit{right}''), and c) maintaining sensitivity to word order changes that could alter the overall semantic composition (such as the difference between sentence with correct order and with shuffling order).
\end{itemize}

This paper is an early step in probing the linguistic knowledge representation in VLP multimodal alignment, covering low-level lexicon, middle-level syntax, and high-level semantic and reasoning knowledge. To our knowledge, our SNARE is the first alignment probing benchmark for VLP, which could snare and reveal the shortages of current VLP models. We hope the probed disadvantages in state-of-the-art VLP models could promote the development of multimodal pretraining.

The remaining sections are organized as follows: 
Sec.~\ref{sec:relate_work} presents related work, introducing existing language knowledge probing and vision-language knowledge probing approaches.  
Sec.~\ref{sec: SNARE} describes the collection and processing of the data and elaborates on the construction of SNARE. 
Sec.~\ref{sec: Experiment} outlines the experimental setup. 
We show the experimental results and suggestions in Sec.~\ref{sec: analyze} and~\ref{subsec: suggestion}, respectively. 
Finally, we conclude in Sec.~\ref{sce:conclusion}.

\begin{table*}[htbp]
\centering
\caption{Based on the semantics level, we compare the existing probing methods by adopting an approximate categorization of their probing targets (e.g., in the FOIL, the probing target is the correspondence between nouns and their modifiers in both the scene and the sentence). The methods marked with \dag are in the form of datasets, while the others are benchmarks. The VL-Checklist does not mention the benchmark size, and we calculate the number of its base datasets as a substitute.}
\label{tab:methods}
\begin{tabular}{lccccccc}
\toprule[1pt]
\multirow{2}{*}{\diagbox{\scriptsize{Methods}}{\begin{tabular}[c]{@{}c@{}}\scriptsize{Relation}\\\scriptsize{Source}\end{tabular}}}  & \footnotesize{Object$\leftrightarrow$Modifier} & \multicolumn{2}{c}{\footnotesize{Object$\leftrightarrow$Object}} & \multicolumn{2}{c}{\footnotesize Sentence$\circlearrowleft$} & \multirow{2}{*}{\footnotesize Base Dataset} & \multirow{2}{*}{Size} \\ \cmidrule(lr){2-2} \cmidrule(lr){3-4} \cmidrule(lr){5-6}
             & \begin{tabular}[c]{@{}c@{}}\scriptsize{Attribute} \\\scriptsize{Ownership}\end{tabular} & \begin{tabular}[c]{@{}c@{}}\scriptsize{Relationship}\\ \scriptsize{Composition}\end{tabular} & \begin{tabular}[c]{@{}c@{}}\scriptsize{Special}\\ \scriptsize{Relationship}\end{tabular} & \begin{tabular}[c]{@{}c@{}}\scriptsize{Semantic}\\ \scriptsize{Structure (Order)}\end{tabular} & \begin{tabular}[c]{@{}c@{}}\scriptsize{Logic}\\ \scriptsize{Relationship}\end{tabular} &                               &                      \\ \midrule[1pt]
FOIL~\textsuperscript{\dag}~\cite{shekhar2017foil}                     & $\checkmark$                   &                          &                      &                            &                    & COCO                      & 123K               \\
\rowcolor{gray!10} SVO-Probes~\textsuperscript{\dag}~\cite{hendricks2021probing}               &                     & $\checkmark$                        &                      &                            &                    & human annotators          & 14K                \\
Salin~\textit{et al.}~\cite{salin2022vision}             & $\checkmark$                   & $\checkmark$                        & $\checkmark$                    &                            &                    &\begin{tabular}[c]{@{}c@{}}\scriptsize{Flickr30k, COCO,}\\ \scriptsize{Flower-102~\cite{Nilsback08}}\end{tabular}  & 6K                 \\
\rowcolor{gray!10} Liu~\textit{et al.}~\cite{liu2022things}               &                     & $\checkmark$                        & $\checkmark$                    &                            &                    & human annotators          & 2K                 \\
VL-Checklist~\cite{zhao2022vl}             & $\checkmark$                   & $\checkmark$                        & $\checkmark$                    &                            &                    &\begin{tabular}[c]{@{}c@{}}\scriptsize{VG, VAW~\cite{pham2021learning},}\\ \scriptsize{HAKE~\cite{li2019hake}, SWiG~\cite{pratt2020grounded}}\end{tabular}         & \textgreater{410K} \\
\rowcolor{gray!10} Winoground~\cite{thrush2022winoground}               & $\checkmark$                   & $\checkmark$                        &                      &                            & $\checkmark$                  & human annotators          & \textless{1K}      \\
ARO~\cite{yuksekgonul2022and}                      & $\checkmark$                   & $\checkmark$                        &                      & $\checkmark$                          &                    & VG, COCO, Flickr30k         & 58K                \\ 
\rowcolor{gray!10} \textbf{SNARE}                    & \textbf{$\checkmark$}                   & \textbf{$\checkmark$}                        & \textbf{$\checkmark$}                    & \textbf{$\checkmark$}                          & \textbf{$\checkmark$} (Negation)       & VG, COCO, Flickr30k         & 76K   \\\bottomrule[1pt]
\end{tabular}
\end{table*}

\section{Related Work}
\label{sec:relate_work}
\subsection{Language Knowledge Probing}
There are a series of studies in the field of natural language processing (NLP) exploring how to probe the linguistic knowledge (e.g., surface-, lexical-, syntactic-, and semantic knowledge) implicitly learned by the neural network models, e.g., language models~\cite{conneau-etal-2018-cram,zhong2022e2s2} and machine translation~\cite{ding2020context,ding2021understanding,zhang2022bliss}.
The studies conducted by Pham~\textit{et al.}~\cite{pham2021out} highlighted that fine-tuning the BERT~\cite{devlin2018bert} on the representative language understanding benchmark -- GLUE~\cite{wang2018glue} may overlook the word order information. 
Sinha~\textit{et al.}~\cite{sinha2021masked} confirmed that word order information is not important during the pretraining of the large language models.
O'Connor~\textit{et al.}~\cite{o2021context} pointed out that for long-range contexts, Transformers~\cite{vaswani2017attention} use co-occurrence statistics of content words to predict the next words. 
Ettinger~\textit{et al.}~\cite{ettinger2020bert} used psychological tasks to evaluate the contextual information of BERT and found that BERT is insensitive to negation factors, a characteristic that we also observe in VLP models. 
Parthasarathi~\textit{et al.}~\cite{parthasarathi2021sometimes} and Sinha~\textit{et al.}~\cite{sinha2021unnatural} studied how models recognize syntax, while Krishna~\textit{et al.}~\cite{krishna2017visual} and Warstadt~\textit{et al.}~\cite{warstadt2020blimp} investigated the complex interaction between syntax and semantic categories in language models. 
In our work, we draw on language evaluation methods, including word shuffling and semantic reversal, to construct the probing benchmark for probing the alignment of multimodal VLP models. We observe similarities between the performance of VLP models and that of the pretrained language models, such as the low sensitivity of multimodal alignment to word order.

\subsection{Vision-Language Knowledge Probing}
Previous studies in the multilingual NLP field have shown that learning accurate alignment (word-, phrase-, and structural-level) between the source-target pairs could bring significantly better source-side language understanding and target-side language generation~\cite{Ding2020SelfAttentionWC,Wu2021MirrorAlignAS}. 
Similarly, one of the keys to cross-modality learning is to develop accurate vision-language knowledge alignment.
To this end, how to appropriately probe such cross-modality alignment becomes important.

This research direction has evolved from studying the mutual interactions between features to the large-scale, rich feature alignment.
Choi~\textit{et al.}~\cite{choi2012context} found that contextual information in images affects the model's understanding of the text.
Cao~\textit{et al.}~\cite{cao2020behind} noted that pretraining models emphasize textual information during inference and that there are potential correspondences between image regions and text words in the attention matrices. 
Frank~\textit{et al.}~\cite{frank2021vision} found that the sharing of information between text and vision is unbalanced, with feature representations of the text encoder being more influenced by visual features. 
Parcalabescu~\textit{et al.}~\cite{parcalabescu2021seeing} found that vision language models have a poor perception of object quantity information in visual input. 
Thrush~\textit{et al.}~\cite{thrush2022winoground} formed a 400-sample Winoground dataset using a manual approach to investigate the perception of features such as objects, actions, and symbolic representations in visual language models. 
Yuksekgonul~\textit{et al.}~\cite{yuksekgonul2022and} developed a large-scale Attribution, Relation, and Order (ARO) benchmark that consists of 50,000 examples designed to evaluate relationships and attributes with fine granularity.

We follow the ARO benchmark and extend it to reflect the semantic and syntax level knowledge required in VLP alignment. 
Therefore, our SNARE benchmark offers challenging probing tasks (with finer-granular options) without sacrificing sample simplicity. 
Apart from feature alignment, we also focus on exploring linguistic knowledge, including semantics, syntax, and so on, to determine how much they impact and enhance VLP alignment.

\begin{table*}[htbp]
\centering
\caption{The average class probability and standard deviation of three random experiments on Semantic Structure.}
\label{tab: Semantic_structure}
\begin{tabular}{lccccccccc}
\toprule[1pt]
 \multirow{2}{*}{Relation} & \multirow{2}{*}{Random} &  \multicolumn{4}{c}{Flickr30k - Semantic Structure}  & \multicolumn{4}{c}{COCO - Semantic Structure}  \\ \cmidrule(lr){3-6} \cmidrule(lr){7-10}
    & & BLIP  &  CLIP  &  Flava  &  X-VLM  & BLIP  &  CLIP  &  Flava  &  X-VLM  \\ \midrule[1pt]
Correct& 25.0  &  26.1\scriptsize{±0.7}  &  \textbf{65.9\scriptsize{±0.3}}  &  20.3\scriptsize{±0.8}  & \textbf{49.6\scriptsize{±1.0}}  & 29.3\scriptsize{±0.4}  &  \textbf{53.9\scriptsize{±0.5}}  &  8.9\scriptsize{±0.1}          & \textbf{42.5\scriptsize{±0.4}} \\
shuffle non-content words & 25.0  & \textbf{43.7\scriptsize{±0.6}}  & 13.0\scriptsize{±0.2}  & 27.7\scriptsize{±0.7}  & 26.0\scriptsize{±1.2}  & \textbf{37.7\scriptsize{±0.6}}  & 16.9\scriptsize{±0.2}  & 29.3\scriptsize{±0.4}  & 28.0\scriptsize{±0.3}          \\
shuffle content words & 25.0  & 18.1\scriptsize{±0.9}  & 15.8\scriptsize{±0.7}  & 19.0\scriptsize{±2.1}  & 14.0\scriptsize{±0.7}  & 18.7\scriptsize{±0.6} & 20.6\scriptsize{±0.3}  & 15.0\scriptsize{±0.5}  & 14.6\scriptsize{±0.3}          \\
shuffle all & 25.0  & 12.1\scriptsize{±0.5}  & 5.4\scriptsize{±0.7}  & \textbf{33.0\scriptsize{±2.0}}  & 10.4\scriptsize{±0.8}  & 14.3\scriptsize{±0.3}  & 8.7\scriptsize{±0.3}  & \textbf{46.7\scriptsize{±0.5}}  & 15.0\scriptsize{±0.2}    \\ \bottomrule[1pt]
\end{tabular}
\end{table*}

\begin{table*}[htbp]
\centering
\caption[\protect]{The class probability on SNARE (Negation Logic, Attribute Ownership, and Relationship Composition). We respectively removed the \textit{Sep} class in Attribute Ownership and the \textit{None} class in Relationship Composition, as reproduction of VG-Attribute and VG-Relationship tasks in the \colorbox{gray!20}{ARO~\cite{yuksekgonul2022and}}.}
\label{tab: SNARE}
\begin{tabular}{ccccc>{\columncolor{gray!20}}cccc>{\columncolor{gray!20}}c}
\toprule[1pt]
\multirow{2}{*}{Models} & Negation Logic  & \multicolumn{3}{c}{Attribution Ownership}& \multirow{2}{*}{VG-Attribution~\cite{yuksekgonul2022and} $\uparrow$}  & \multicolumn{3}{c}{Relationship Composition} & \multirow{2}{*}{VG-Relation~\cite{yuksekgonul2022and} $\uparrow$} \\ \cmidrule(lr){3-5} \cmidrule(lr){7-9} 
                        &\textit{Cor} $\uparrow$  &\textit{Cor} $\uparrow$       & \textit{Sep} $\uparrow$     & \textit{Exc} $\downarrow$  &   & \textit{Cor} $\uparrow$        & \textit{Exc} $\downarrow$        & \textit{None}$ \uparrow$   &                        \\ \midrule[1pt]
Random &50.0  &33.3&33.3&33.3&50.0&33.3&33.3&33.3&50.0   \\ \midrule[0.5pt]
BLIP                    & 79.0        & 48.9        & 45.2         & 5.9      & 85.3   & 41.2           & 35.1            & 23.7       &  54.6                   \\
CLIP                    & 47.3        & 40.0        & 36.1         & 23.9     & 61.7   & 38.3           & 36.8            & 25.0       &  51.7                   \\
FLAVA                   & 12.9        & 67.9        & 1.6          & 30.5     & 68.7   & 0.8            & 1.9             & 97.3       &  43.6                   \\
X-VLM                   & 48.1        & 54.0        & 39.1         & 6.9      & 85.6   & 30.0           & 23.9            & 46.0       &  57.1                   \\ \bottomrule[1pt]
\end{tabular}
\end{table*}

\section{SNARE Benchmark Construction}
\label{sec: SNARE}
Previous probing benchmarks~\cite{salin2022vision, thrush2022winoground, shekhar2017foil, zhao2022vl} provide two options in one sample to assess the effectiveness of semantic level multimodal alignment and usually overlook syntactic level. 
Following the multi-choice question-answering approach from Li~\textit{et al.}~\cite{li2023seed}, we devise more options in tasks of SNARE to explore the impact of syntactic- and semantic-level alignment~(\S\ref{subsubsec: SS},~\ref{subsubsec: AO},~\ref{subsubsec: SR}). 
In addition, considering the significance of reasoning~\cite{zeijlstra2007negation, orenes2014negation}, we present the Negation Logic task~(\S\ref{subsubsec: NL}).

Firstly, we introduce how we obtain and process the data.
Then, we explain why and how to structure the four tasks (\textit{B}. Semantic Structure, \textit{C}. Negation Logic, \textit{E}. Attribute Ownership, and \textit{D}. Relationship Composition) in the SNARE benchmark. The detailed samples are shown in Fig.~\ref{fig:overview} to facilitate understanding.

\subsection{Data Collection}
\label{subsec: data}
In the SNARE benchmark, fine-grained image-text features are required. 
We obtain explicit features from three commonly-used high-quality multimodal datasets, including Visual Genome~\cite{krishna2017visual}, COCO~\cite{lin2014microsoft}, and Flickr30k~\cite{young2014image}, and process them through the method proposed by Yuksekgonul~\textit{et al.}~\cite{yuksekgonul2022and}, and the Spacy toolkit~\cite{honnibal2017spacy}.

To process \textbf{the Visual Genome dataset}, we extract explicit visual and textual features through the following steps:
\textit{\textbf{1)}} traversing through the scene graphs annotated in GQA~\cite{hudson2019gqa} and identifying the objects with bounding box;
\textit{\textbf{2)}} to ensure salience, discarding objects whose bounding box's width or height is less than 1/4 of that of the image;
\textit{\textbf{3)}} randomly selecting two different objects $\boldsymbol{X}, \boldsymbol{Y}$, where $\boldsymbol{X}=\boldsymbol{Y}$, and extracting their corresponding attribute $\boldsymbol{A}, \boldsymbol{B}$ or the space/verb relationships~$\boldsymbol{R}$ between them, where $\boldsymbol{X}$ is the subject and $\boldsymbol{Y}$ is object;
\textit{\textbf{5)}} to reduce interference from the rich visual content, extracting a minimal bounding box containing $\boldsymbol{X}$ and $\boldsymbol{Y}$ from the scene as image $\boldsymbol{I}$;
\textit{\textbf{6)}} finally, obtaining \textbf{the nouns-relation dataset}, whose samples contain features~$\{\boldsymbol{I}, \boldsymbol{X}, \boldsymbol{Y}, \boldsymbol{R}\}$, and \textbf{the nouns-attributes dataset}, whose samples contain features~$\{\boldsymbol{I}, \boldsymbol{X}, \boldsymbol{Y}, \boldsymbol{A}, \boldsymbol{B}\}$.

To tag the part of speech $a_i$ of words $t_i$ in the text $\boldsymbol{T}$ from \textbf{the COCO and Flickr30k dataset}, we employ the Spacy~\cite{honnibal2017spacy} toolkit to parse the sentences. 
We obtain a set of sample features $\{\boldsymbol{I}, \boldsymbol{T}, a_1, \dots, a_l\}$, where $l$ represents the text length and $a_i \in \boldsymbol{S}_{pos}$. $\boldsymbol{S}_{pos}=\{noun, adjective, verb, \dots\}$ is a set including all POS tags\footnote{For additional detailed information on the POS tags of the Spacy toolkit, please refer to https://universaldependencies.org/u/pos/.} in the Spacy toolkit. 

\subsection{Dataset for Semantic Structure}
\label{subsubsec: SS} 
The Semantic Structure\footnote{``Semantic Structures'' means the conceptual structure of the sentence and its lexical and syntactic expression~\cite{jackendoff1992semantic}. Shuffling word order destroys all structure in the sentence and we use ``Semantic Structure'' as the task name.} task aims to investigate whether the syntax structure (words order) and semantics composition (combination between different parts of speech) influence the alignment. 
It is constructed by disrupting the syntax order and retaining discrete semantics. We re-organize the processed COCO and Flickr30k data features $\{\boldsymbol{I}, \boldsymbol{T}, a_1, . . . , a_l\}$. 
We define a part of speech set $\boldsymbol{E} \subseteq \boldsymbol{S}_{pos}$ and mark $t_i$ whenever $a_i \in \boldsymbol{E}$. The function $\boldsymbol{f}(\boldsymbol{T}, a_1, . . . , a_l, \boldsymbol{E})$ is defined to shuffle the marked $t_i$ in the text sequence $\boldsymbol{T}$. The sample representation is obtained as follows:

\begin{equation}
    \boldsymbol{I} \text{ and } \boldsymbol{f}(\boldsymbol{T}, a_i|_{1}^l, \boldsymbol{E})
    \begin{cases}
        \text{\textit{Origin}:} & \boldsymbol{E} = \emptyset\\
        \text{\textit{\footnotesize Shuffle content words}:} & \boldsymbol{E} = \boldsymbol{C}\\
        \text{\textit{\footnotesize Shuffle non-content words}:} & \boldsymbol{E} = \boldsymbol{\overline{C}} \\
        \text{\textit{Shuffle all}:} & \boldsymbol{E} = \boldsymbol{S}_{pos}
    \end{cases},
\end{equation}
where $\boldsymbol{C} = \{noun, adjective, verb\}$ represents the type set of content words\footnote{Content words, in linguistics, are words that possess semantic content and contribute to the meaning of the sentence in which they occur. We simply selected nouns, adjectives, and verbs that prominently represent visual semantics as content words.}. 
For instance, we provide an example in the ``Semantic Structure'' part in Fig.~\ref{fig:overview}. 
The ``Origin'' is ``\textit{\textcolor{blue}{a$_{\_determiner}$} \textcolor{green}{girl$_{\_noun}$} \textcolor{green}{sits$_{\_verb}$} \textcolor{blue}{on$_{\_determiner}$} \textcolor{blue}{a$_{\_determiner}$} \textcolor{green}{decorated$_{\_verb}$} \textcolor{green}{bike$_{\_noun}$} \textcolor{blue}{with$_{\_adposition}$} \textcolor{blue}{a$_{\_determiner}$} \textcolor{green}{younger$_{\_adjective}$} \textcolor{green}{boy$_{\_noun}$} \textcolor{blue}{while$_{\_subordinating \; conjunction}$} \textcolor{blue}{another$_{\_determiner}$} \textcolor{green}{girl$_{\_determiner}$} \textcolor{green}{takes$_{\_verb}$} \textcolor{blue}{a$_{\_determiner}$} \textcolor{green}{picture$_{\_noun}$}}'', which is a reference sentence $T$ from COCO or Flickr30k without shuffling. The subscripts indicate part of speech $a_i$. 
Words are annotated in green whose \textcolor{green}{$a_i \in \boldsymbol{C}$}, and the others whose \textcolor{blue}{$a_i \in \boldsymbol{\overline{C}}$} are annotated in blue. 
In ``shuffle content words'', we shuffle the green part in $\boldsymbol{T}$, like ``\textit{a \textcolor{green}{girl (girl) boy (sit)} on a \textcolor{green}{bike (decorated) picture (bike)} with a \textcolor{green}{younger (younger) sits (boy)} while another \textcolor{green}{decorated (girl) takes (takes)} a \textcolor{green}{girl (picture)}}'', where the original words are provided in parentheses. 
Similarly, in ``shuffle non-content words'' and ``shuffle all'', we shuffle the blue part and all words in $\boldsymbol{T}$ respectively.

\subsection{Dataset for Negation Logic}
\label{subsubsec: NL}
Humans can infer missing visual pieces by understanding negation commands, such as ``Not'' and ``No'', in multimodal information processing~\cite{zeijlstra2007negation, orenes2014negation}. By introducing the syntax rules, the Negation Logic task incorporates the negation word (``\textit{not}'') into the nouns-attributes dataset to evaluate logical reasoning abilities. 

In Negation Logic, each sample comprises an image, a positive statement (\textit{Correct} class, \textit{Cor}), and a negative statement (\textit{Wrong} class, \textit{Wro}). The sample structure is illustrated below:

\begin{equation}
    \boldsymbol{I} \text{ and } 
    \begin{cases}
    \text{\textit{Cor}:} & \text{the}\; \boldsymbol{X}\; \text{is}\; \boldsymbol{A}\; \text{and the}\; \boldsymbol{Y}\; \text{is}\; \boldsymbol{B} \\
    \text{\textit{Wro}:} & \text{the}\; \boldsymbol{X}\; \text{is \textbf{not}}\; \boldsymbol{A}\; \text{and the}\; \boldsymbol{Y}\; \text{is \textbf{not}}\; \boldsymbol{B}\; 
\end{cases}
\end{equation}

We show an example in Fig.\ref{fig:overview} in ``Negation Logic'' part, where the \textit{Cor} sentence is ``\textit{the bus ($\boldsymbol{X}$) is white ($\boldsymbol{A}$) and  the road ($\boldsymbol{Y}$) is black ($\boldsymbol{B}$)}'' and the \textit{Wro} sentence is ``\textit{the bus ($\boldsymbol{X}$) is \textbf{not} white ($\boldsymbol{A}$) and the road ($\boldsymbol{Y}$) is \textbf{not} black ($\boldsymbol{B}$)}''.

\subsection{Dataset for Attribute Ownership}
\label{subsubsec: AO}
Humans construct complex semantics by combining adjectives and nouns utilizing various syntax forms~\cite{fyshe2019lexical}, enabling them to envision elaborate visual scenes~\cite{hsieh2023sugarcrepe}. 
The Attribute Ownership task aims to assess the VLP models' ability to the semantic match between vision and language and understanding of syntax (short- and long-distance). 
For two sentences conveying the same semantic meaning, we label the sentence, in which nouns have a shorter distance to adjectives than the other, as a short-distance syntax sentence, and the other as a long-distance syntax sentence.
Each sample in Attribute Ownership includes a short- (\textit{Correct} class, \textit{Cor}) and a long-distance (\textit{Separate} class, \textit{Sep}) syntax sentence, and a sentence mismatching with $\boldsymbol{I}$ (\textit{Exchange} class, \textit{Exc}).

We construct samples by utilizing the nouns-attributes dataset. 
In \textit{Cor}, noun and attribute pairs are closely connected (with distance 1), and easy to merge semantics, like ``\textit{the blue ($\boldsymbol{A}$) sky ($\boldsymbol{X}$) and the large ($\boldsymbol{B}$) building ($\boldsymbol{Y}$)}''. 
The \textit{Sep} has a longer semantic expression (with distance 5/ 6), like ``\textit{the sky ($\boldsymbol{X}$) and the building ($\boldsymbol{Y}$) are blue ($\boldsymbol{A}$) and large ($\boldsymbol{B}$) respectively}''. 
It is less frequently used but still enables humans to easily understand the image content. 
The \textit{Exc} swaps the position of attributes and nouns in \textit{Cor}, resulting in a mismatch semantic with the image, like ``the large ($\boldsymbol{B}$) sky ($\boldsymbol{X}$) and the blue ($\boldsymbol{A}$) building ($\boldsymbol{Y}$)''. 
Three classes in samples are illustrated below:

\begin{equation}
    \boldsymbol{I} \text{ and } 
    \begin{cases}
    \text{\textit{Cor}:} & \text{the}\; \boldsymbol{A}\; \boldsymbol{X}\; \text{and the}\; \boldsymbol{B}\; \boldsymbol{Y}\\
    \text{\textit{Sep}:} & \text{the}\; \boldsymbol{X}\; \text{and the}\; \boldsymbol{Y}\; \text{are}\; \boldsymbol{A}\; \text{and}\; \boldsymbol{B}\; \text{respectively}\\
    \text{\textit{Exc}:} & \text{the}\; \boldsymbol{B}\; \boldsymbol{X}\; \text{and the}\; \boldsymbol{A}\; \boldsymbol{Y}\;
\end{cases}
\end{equation}

\begin{figure*}[htbp]
  \centering
  \subfigure[BLIP (Acc=31.4\%)]{\label{fig:Spatial_Relation-blip}\includegraphics[width=0.24\textwidth]{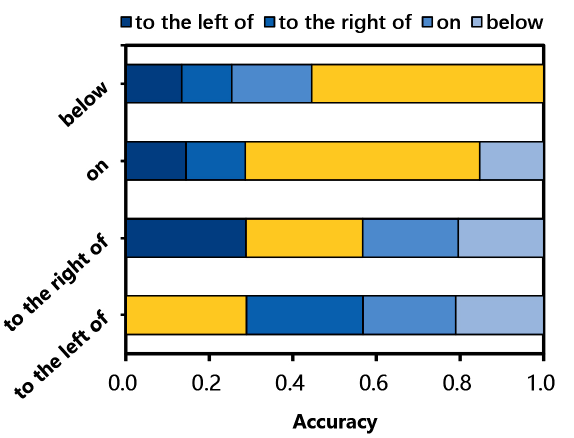}}
  \subfigure[CLIP (Acc=40.8\%)]{\label{fig:Spatial_Relation-clip}\includegraphics[width=0.24\textwidth]{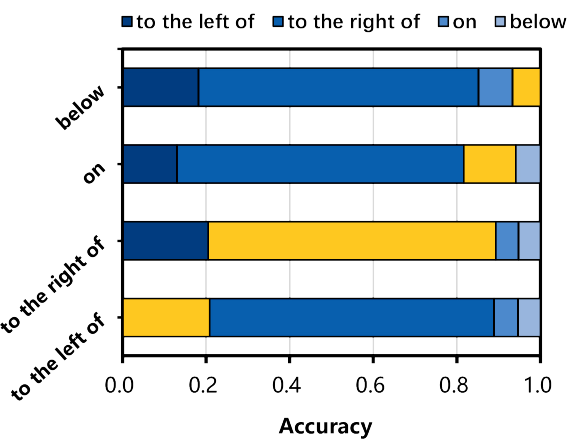}}
  \subfigure[Flava (Acc=10.8\%)]{\label{fig:Spatial_Relation-flava}\includegraphics[width=0.24\textwidth]{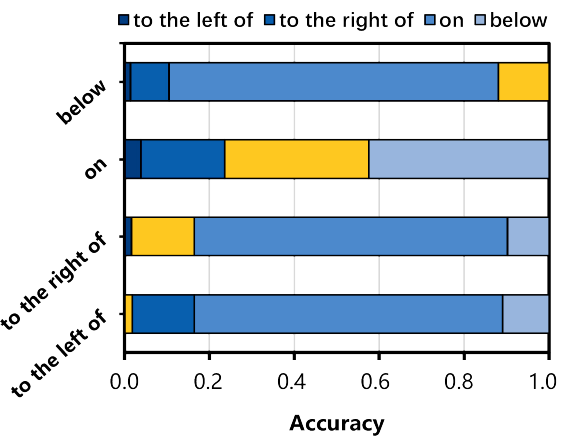}}
  \subfigure[X-VLM ((Acc=34.1\%)]{\label{fig:Spatial_Relation-x-vlm}\includegraphics[width=0.24\textwidth]{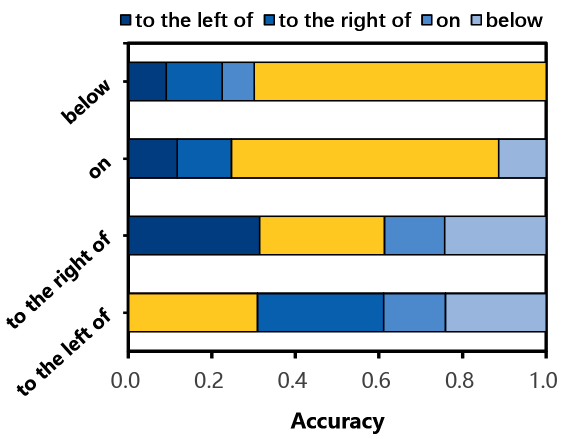}}
  
  \caption{Performance on the Multi-spatial Relationship. The vertical axis indicates the correct spatial relations in the sentence. The legend indicates the class. CLIP and Flava display a preference towards specific spatial relationships (``to the right of'' and ``on''). 
  }
  \label{fig: Spatial_Relation}
\end{figure*}

\subsection{Dataset for Relationship Composition}
\label{subsubsec: SR}
The Relationship Composition assesses whether the VLP model can accurately comprehend the relationship between two objects and whether it is sensitive to more multi-element (two or three) word combinations. 

We combine features from the nouns-relation dataset and devise three different sentences for each sample. 
The \textit{Correct} class sentence (\textit{Cor}) describes the relationship composition between $\boldsymbol{X}$ and $\boldsymbol{Y}$ accurately (triplet. $\boldsymbol{X}$ is the initiator of $\boldsymbol{R}$ and $\boldsymbol{Y}$ is the recipient, for example, ``\textit{the girl~($\boldsymbol{X}$) is wearing~($\boldsymbol{R}$) the shirt~($\boldsymbol{Y}$)}''). 
In the \textit{Exchange} class sentence (\textit{Exc}), the position between $\boldsymbol{X}$ and $\boldsymbol{Y}$ is exchanged, like ``\textit{the shirt~($\boldsymbol{Y}$) is wearing~($\boldsymbol{R})$ the girl~($\boldsymbol{X}$)}''. 
The \textit{None} class (\textit{None}) removes the relationship word $\boldsymbol{R}$ and becomes a binary tuple comprised of nouns, like ``\textit{the girl ($\boldsymbol{X}$) and the shirt ($\boldsymbol{Y}$)}''. 
The sample structure is as follows:

\begin{equation}
    \boldsymbol{I} \text{ and } 
    \begin{cases}
    \text{ \textit{Cor}:} & \text{the}\; \boldsymbol{X}\; \text{is}\; \boldsymbol{R}\; \text{the}\; \boldsymbol{Y} \\
        \text{ \textit{Exc}:} & \text{the}\; \boldsymbol{Y}\; \text{is}\; \boldsymbol{R}\; \text{the}\; \boldsymbol{X}\;\\
    \text{\textit{None}:} & \text{the}\; \boldsymbol{X}\; \text{and}\; \text{the}\; \boldsymbol{Y}
\end{cases}
\end{equation}

We develop a sub-task from Relationship Composition, named the \textit{Multi-spatial Relationship}. It focuses on evaluating the ability to distinguish the spatial relationship in vision.
we filter data from the nouns-relation dataset when
\begin{equation}
\boldsymbol{R} \in \{\text{``\textit{to the left of}”}, \text{``\textit{to the right of}”}, \text{``\textit{on}”}, \text{``\textit{below}”}\}, 
\end{equation}
representing four different direction relationships. 
The sample structure is as follows:

\begin{equation}
    \boldsymbol{I} \text{ and } 
    \begin{cases}
    \text{the}\; \boldsymbol{X}\; \text{is \textit{to the left of} the}\; \boldsymbol{Y}\\
    \text{the}\; \boldsymbol{X}\; \text{is \textit{to the right of} the}\; \boldsymbol{Y}\\
    \text{the}\; \boldsymbol{X}\; \text{is \textit{on} the}\; \boldsymbol{Y}\\
    \text{the}\; \boldsymbol{X}\; \text{is \textit{below} the}\; \boldsymbol{Y}\\
\end{cases},
\end{equation}
where only the sentence with the correct spatial relationship $\boldsymbol{R}$ which matches with the $\boldsymbol{I}$ is the right class.

\begin{table*}[htbp]
\centering
\caption{Performance on each sub-category in Relationship Composition. 
When models have low performance in distinguishing relationships (\colorbox{red!10}{\textit{Cor} - \textit{Exc}$<5\%$}, no statistically significant difference), we highlight the corresponding scores. When there are three models achieving low performance on the same sub-category, we highlight the \colorbox{red!20}{category} name. 
We only show  spatial-based Relationships whose occurrences are larger than 25 and verb-based ones whose occurrences are larger than 20.}
\label{tab: Relationship Composition}
\begin{tabular}{lccccccccccccr}
\toprule[1pt]
\multirow{2}{*}{Relation} & \multicolumn{3}{c}{BLIP}  & \multicolumn{3}{c}{CLIP}  & \multicolumn{3}{c}{Flava}  & \multicolumn{3}{c}{X-VLM} & \multirow{2}{*}{Freq} \\ \cmidrule(lr){2-4}  \cmidrule(lr){5-7} \cmidrule(lr){8-10} \cmidrule(lr){11-13}
                          & \textit{Cor}$\uparrow$ & \textit{Exc}$\downarrow$ & \textit{None}$\uparrow$ & \textit{Cor}$\uparrow$ & \textit{Exc}$\downarrow$ & \textit{None}$\uparrow$ & \textit{Cor}$\uparrow$ & \textit{Exc}$\downarrow$ & \textit{None}$\uparrow$  & \textit{Cor}$\uparrow$ & \textit{Exc}$\downarrow$ & \textit{None}$\uparrow$ &                       \\ \midrule[1pt]
\multicolumn{14}{c}{Spatial-based Relationship}     \\ 
Acc($25<$Freq$<5112$)             & \textbf{52.9}    & 22.3     & 24.8 & 28.5    & 26.1     & \textbf{45.4} & 0.6     & 5.4      & \textbf{94.0}  & 41.0    & 13.1      & \textbf{45.9} & 4865  \\ 
\hline
above                     & 43.5    & 26.8     & 29.7 & \cellcolor{red!10}28.3    & \cellcolor{red!10}33.5     &38.3 & 0.0     & 0.4      & 99.6  & 41.3    & 6.3      & 52.4 & 269                   \\
at                        & 54.7    &32.0     &13.3 & 36.0    & 21.3     & 42.7 & 2.7     & 6.7      & 90.7  & \cellcolor{red!10}29.3    & \cellcolor{red!10}30.7     & 40.0 & 75                    \\
behind                    & 56.8    & 16.4     & 26.8 & 29.8    & 24.0     & 46.2 & 0.7     & 0.9      & 98.4  & 46.3    & 10.6     & 43.0 & 574                   \\
below     & 49.3    & 25.4     & 25.4 & 34.9    & 23.4     & 41.6 & 1.0     & 0.0      & 99.0  & 48.8    & 13.9     & 37.3 & 209                   \\
in                        & 58.8    & 18.9     & 22.3 & 30.6    & 18.5     & 50.8 & 0.7     & 14.8     & 84.5  & 32.3    & 17.9     & 49.7 & 708                   \\
in front of               & 50.3    & 36.4     & 13.3 & \cellcolor{red!10}32.5    & \cellcolor{red!10}27.9     & 39.6 & 0.3     & 0.0      & 99.7  & 46.4    & 22.3     & 31.3 & 588                   \\
inside                    & 56.9    & 29.3     & 13.8 & \cellcolor{red!10}34.5    & \cellcolor{red!10}36.2     & 29.3 & 0.0     & 5.2      & 94.8  & 46.6    & 20.7     & 32.8 & 58                    \\
of                        & 42.2    & 23.4     & 34.3 & \cellcolor{red!10}28.9    & \cellcolor{red!10}41.4     & 29.7 & 1.1     & 12.3     & 86.6  & \cellcolor{red!10}30.2    & \cellcolor{red!10}26.2     & 43.6 & 367                   \\
on                        & 54.8    & 19.4     & 25.9 & \cellcolor{red!10}24.9    & \cellcolor{red!10}25.4     & 49.7 & 0.4     & 5.7      & 93.9  & 41.7    & 6.1      & 52.1 & 1684                  \\
on top of                 & 50.7    & 18.4     & 30.8 & \cellcolor{red!10}21.9    & \cellcolor{red!10}29.4     & 48.8 & 0.5     & 2.0      & 97.5  & 44.8    & 9.5      & 45.8 & 201                   \\
\cellcolor{red!20}to the left of            & \cellcolor{red!10}37.7    & \cellcolor{red!10}38.5     & 23.8 & \cellcolor{red!10}41.5    & \cellcolor{red!10}41.0     & 17.5 & 1.0     & 0.9      & 98.1  & \cellcolor{red!10}26.5    & \cellcolor{red!10}27.1     & 46.4 & 7741(-)                  \\
\cellcolor{red!20}to the right of           & \cellcolor{red!10}36.2    & \cellcolor{red!10}38.5     &  25.3 & \cellcolor{red!10}44.0    & \cellcolor{red!10}44.6     &11.4 & 1.1     & 1.1      & 97.8  & \cellcolor{red!10}25.2    & \cellcolor{red!10}28.9     & 45.9 & 7741(-)                  \\
under                     & 49.2    & 19.7     & 31.1 & 33.3    & 16.7     & 50.0 & 0.8     & 0.8      & 98.5  & 45.5    & 13.6     & 40.9 & 132                   \\ \hline
\multicolumn{14}{c}{Verb-based Relationship}               \\ 
Acc($25<$Freq$<425$)                & \textbf{52.5}    & 31.1     & 16.4 & 36.0    & 21.9     & \textbf{42.0} & 0.3     & 1.3      & \textbf{98.4}  & \textbf{43.9}   & 12.6     & 43.5  &  752  \\ \hline
covered by                & 38.9    & 30.6     & 30.6 & \cellcolor{red!10}13.9    & \cellcolor{red!10}30.6     & 55.6 & 2.8     & 0.0      & 97.2  & \cellcolor{red!10}33.3    & \cellcolor{red!10}36.1     & 30.6 & 36                    \\
covering                  & 45.5    & 18.2     & 36.4 & \cellcolor{red!10}15.2    & \cellcolor{red!10}27.3     & 57.6 & 0.0     & 0.0      & 100.0 & \cellcolor{red!10}27.3    & \cellcolor{red!10}42.4     & 30.3 & 33                    \\
eating                    & 61.9    & 33.3     & 4.8  & 57.1    & 9.5      & 33.3 & 0.0     & 0.0      & 100.0 & 28.6    & 9.5      & 61.9 & 21                    \\
holding                   & \cellcolor{red!10}34.5    & \cellcolor{red!10}51.4     & 14.1 & 28.9    & 18.3     & 52.8 & 0.0     & 0.0      & 100.0 & 22.5    & 14.8     & 62.7 & 142                   \\
looking at                & 45.2    & 25.8     & 29.0 & 38.7    & 3.2      & 58.1 & 0.0     & 0.0      & 100.0 & 29.0    & 0.0      & 71.0 & 31                    \\
lying on                  & 56.7    & 26.7     & 16.7 & 38.3    & 30.0     & 31.7 & 0.0     & 0.0      & 100.0 & 41.7    & 3.3      & 55.0 & 60                    \\
parked on                 & 57.1    & 14.3     & 28.6 & 57.1    & 19.0     & 23.8 & 0.0     & 19.0     & 81.0  & 57.1    & 4.8      & 38.1 & 21                    \\
riding                    & \cellcolor{red!10}43.1    & \cellcolor{red!10}51.0     & 5.9  & 58.8    & 21.6     & 19.6 & 0.0     & 0.0      & 100.0 & 56.9    & 21.6     & 21.6 & 51                    \\
sitting at                & 61.5    & 38.5     & 0.0  & \cellcolor{red!10}34.6    & \cellcolor{red!10}34.6     & 30.8 & 0.0     & 0.0      & 100.0 & 50.0    & 23.1     & 26.9 & 26                    \\
sitting in                & 65.2    & 30.4     & 4.3  & \cellcolor{red!10}26.1    & \cellcolor{red!10}21.7     & 52.2 & 0.0     & 0.0      & 100.0 & 39.1    & 13.0     & 47.8 & 23                    \\
sitting on                & 61.1    & 24.6     & 14.3 & 36.0    & 20.6     & 43.4 & 0.0     & 2.9      & 97.1  & 63.4    & 2.9      & 33.7 & 175                   \\
standing in               & 71.2    & 15.3     & 13.6 & 45.8    & 15.3     & 39.0 & 1.7     & 0.0      & 98.3  & 35.6    & 18.6     & 45.8 & 59                    \\
standing on               & 63.5    & 11.5     & 25.0 & \cellcolor{red!10}38.5    & \cellcolor{red!10}34.6     & 26.9 & 0.0     & 1.9      & 98.1  & 69.2    & 5.8      & 25.0 & 52                    \\
watching                  & \cellcolor{red!10}40.9    & \cellcolor{red!10}40.9     & 18.2 & \cellcolor{red!10}27.3    & \cellcolor{red!10}27.3     & 45.5 & 0.0     & 0.0      & 100.0 & 27.3    & 13.6     & 59.1 & 22                    \\
wearing                   & \cellcolor{red!10}41.5    & \cellcolor{red!10}47.1     & 11.4 & 31.1    & 25.6     & 43.3 & 0.4     & 0.1      & 99.5  & \cellcolor{red!10}24.9    & \cellcolor{red!10}21.7     & 53.4 & 949 (-) \\ \bottomrule[1pt]                 
\end{tabular}
\end{table*}

\begin{algorithm}[htbp]
\label{alg:score}
\caption{Scoring Method for MLLMs}
\begin{algorithmic}[1]
\Require $\{I, T\}$: image-text pair; $F_{m}$: BLIP2 model; $T_p$: prompt $=\text{''Describe the image.''}$; $F_{swi}$: switch sentence to yes/ no questions.
\Ensure \textit{score}
\Function{ScoringMethod}{$\{I, T\}$}
    \State $T_q \gets F_{swi}(T)$
    \State $T_{des} \gets F_{m}(I, \text{``Question:}\{T_p\}\text{ Answer:''})$
    \State $T_{out} \gets F_{m}(I,$$\text{``Question:}\{T_p\}\text{ Answer:}\{T_{des}\}$
    \State $\text{\hspace{100pt} Question:}\{T_q\}\text{ Answer:''})$
    \If{``no'' in $T_{out}$}
        \State $score \gets 0$
    \Else
        \State $score \gets 1$
    \EndIf
    \State \textbf{return} $score$
\EndFunction
\end{algorithmic}
\end{algorithm}

\section{Experiment Setup}
\label{sec: Experiment}
\subsection{Models}
We use the state-of-the-art VLP models from the past two years including \textbf{BLIP} (``base''\footnote{\url{https://github.com/salesforce/BLIP}} variant), \textbf{CLIP} (``VIT-B/32''\footnote{\url{https://github.com/openai/CLIP}} variant), \textbf{Flava} (``flava-full''\footnote{\url{https://huggingface.co/facebook/flava-full}} variant), \textbf{X-VLM} (``base''\footnote{\url{https://github.com/zengyan-97/X-VLM}} variant) and \textbf{BLIP2} (``blip2\_t5''\footnote{\url{https://github.com/salesforce/LAVIS/tree/main/projects/blip2}} variant). The experiment code is modified from Yuksekgonul~\textit{et al.}~\cite{yuksekgonul2022and}\footnote{\url{https://github.com/mertyg/when-and-why-vlms-bow}}, with input text lengths limited to 30. 
For part-of-speech tagging, we employed the ``en\_core\_web\_sm'' annotation model from the Spacy library~\cite{honnibal2017spacy}.

We focus on the linguistic knowledge acquired from VLP.
Hence, we have not fine-tuned any models on downstream tasks.
We calculate the multimodal similarity scores of BLIP and X-VLM using the fully connected layer obtained during model pretraining. The CLIP model calculated the cosine distance between visual and text features as the similarity scores. Flava computes them with the fusion module and the fully connected layer classifier.

\subsection{Evaluation Metrics}
\subsubsection{Accuracy for VLP Models}
The four tasks in SNARE require the VLP model to rank the image-text matching confidence for each sentence. We report the \textbf{per-class accuracy} score on each class. For the Semantic Structure task, we conduct repeated experiments using three different random seeds and present the average and standard deviation as the outcome.

\subsubsection{Binary Classification Accuracy for MLLMs}
The evaluation of MLLMs models is still incomplete~\cite{zhao2023survey}. Taking inspiration from MME~\cite{fu2023mme}, we develop a scoring method that utilizes simple prompts to guide the model's output, shown as a binary classification-like Alg.~\ref{alg:score}. \textbf{Binary classification accuracy} scores for the same class of sentences are used as the evaluation criteria. The scoring is based on the presence of ``\textit{yes}'' or ``\textit{no}'' in the model's answers. 
It is important to note that there are occasional instances where these words are absent in the model's responses, which are regarded as negative examples. In cases where the sentence $T$ in text-image pairs does not contain a copula verb, we transform it into a yes/ no question using the structure ``\textit{are there + $T$ in the image}''.

\begin{figure}[htbp]
\centering
  \subfigure[\textit{Cor} $\uparrow$]{\label{fig:blip2_negative_correct}\includegraphics[scale=0.40]{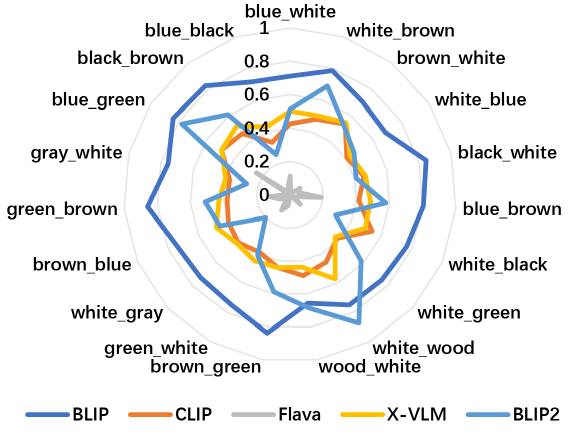}}
  \subfigure[\textit{Wro} $\downarrow$]{\label{fig:blip2_negative_wrong}\includegraphics[scale=0.40]{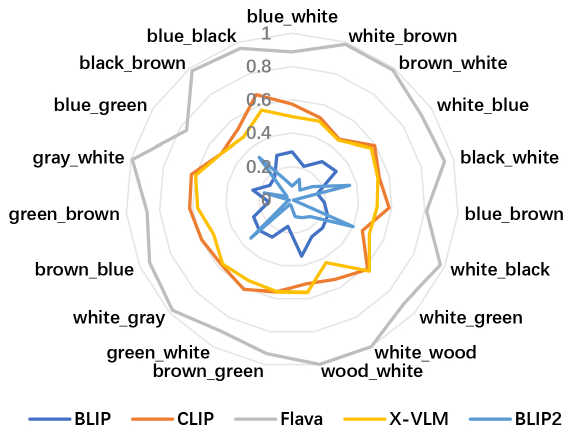}}
\caption{\textbf{The BLIP2 performance on the Negation Logic}. For BLIP, CLIP, Flava, and X-VLM, we present the sub-classification accuracy of the corresponding class. Compared to models that do not utilize LLM as the main component, BLIP2 performs better in understanding negation content.}\label{fig:blip2_negative}
\end{figure}

\begin{figure}[htbp]
\centering
\subfigure[\textit{Cor} $\uparrow$]{\label{fig:blip2_attribute_relationship_correct}\includegraphics[scale=0.4]{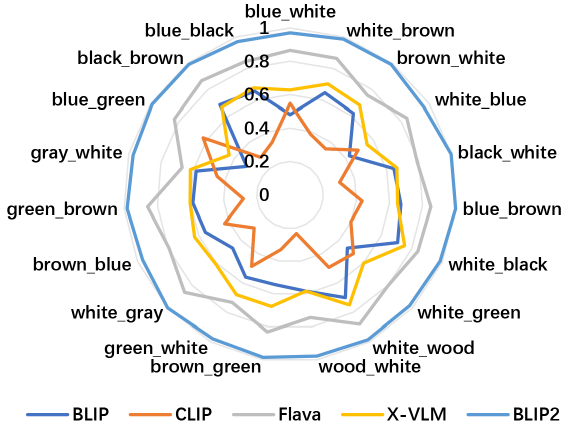}}
\subfigure[\textit{Sep} $\uparrow$]{\label{fig:blip2_attribute_relationship_sparate}\includegraphics[scale=0.4]{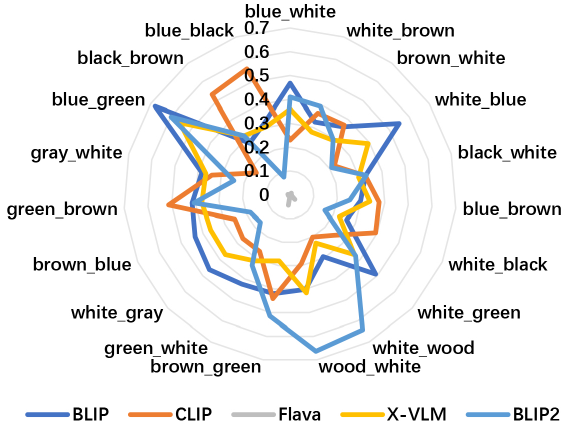}}
\subfigure[\textit{Exc} $\downarrow$]{\label{fig:blip2_attribute_relationship_exchange}\includegraphics[scale=0.4]{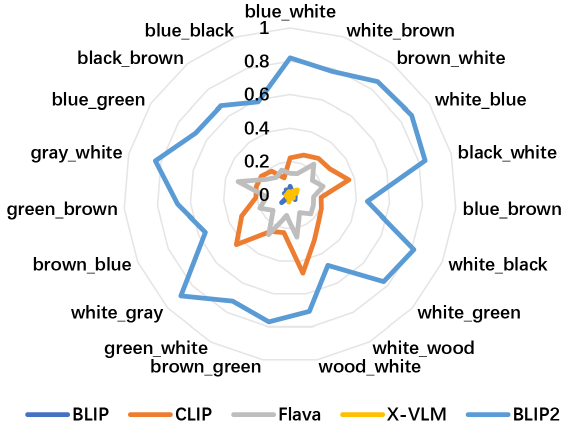}}

\caption{\textbf{The BLIP2 performance on the Attribute Ownership}. BLIP, CLIP, Flava, and X-VLM show the sub-classification accuracy of the corresponding class (``\textit{Cor}'', ``\textit{Sep}'', and ``\textit{Exc}'') in the task. 
}\label{fig:blip2_attribute_relationshi}
\end{figure}

\section{Result and Analysis}
\label{sec: analyze}
We analyze the performance of the VLP model on SNARE across three levels: lexical (\S\ref{subsubsec: lexical}), syntactic (\S\ref{subsubsec: syntactic}), and semantic (\S\ref{subsubsec: semantic}). Separately, we analyze the performance of the MLLMs model (\S\ref{subsubsec: MLLMs}), a new paradigm of VLP models.

\subsection{Lexical Level Probing Results}
\label{subsubsec: lexical}
\textbf{
The content words contain effective words semantic information in constructing sentence semantics, and functional words play a less significant role.}

Tab.~\ref{tab: Semantic_structure} presents the model performance on the Semantic Structure. 
CLIP (65.9\% and 53.9\%) and X-VLM (49.6\% and 42.5\%) exhibit a higher probability of selecting ``correct'', in which content and function words play significant roles in composing sentence semantics.
BLIP displays a higher probability (43.7\% and 37.7\%) to select ``shuffle non-content words'' where the order of function words is shuffled while that of the content words is retained.
It indicates that BLIP prefers content  contributing to regularization.
Flava has a similar performance to BLIP (27.7\% and 29.3\% on ``shuffle non-content words'' $>$ 20.3\% and 8.9\% on ``correct''). However, it is prone to selecting thoroughly unordered sentences (33.0\% and 46.7\% on ``shuffle all''). 
This suggests Flava lacks an advantage in assembling discrete words into more intricate semantic expressions. 
Notably, when the order of content words is disrupted (``shuffle content words'' and ``shuffle all''), all models except Flava exhibit lower selection probabilities. 

This observation underscores the crucial role of content words in conveying sentence semantics regarding part of speech. In contrast, the significance of function words remains relatively secondary, playing a role similar to that of a regularization term.
In terms of word order, there exist models that prefer discrete semantics on the word level (Flava), and most models rely on word order to capture complex and accurate semantics (BLIP, CLIP, and X-VLM).
This does not align with previous research~\cite{yuksekgonul2022and}. VLP models do not always behave like bag-of-words models, and word order is necessary for structuring complex semantics using discrete content words.

\subsection{Syntactic Level Probing Results}
\label{subsubsec: syntactic}
\textbf{
The VLP models are inclined to deal with comprehending short-distance syntactic combinations and simple syntactic relationships.} 

In Tab.~\ref{tab: SNARE}, we present the models' performance on Attribute Ownership and Relationship Composition evaluations. 
In Attribution Ownership, all models tend to choose sentences with short-distance syntactic expressions (\textit{Cor}). 
Flava has a particularly remarkable performance (67.9\%), and it shows a very low probability of selecting long-distance syntactic (1.6\% in \textit{Sep}). 
In Relationship Composition, both Flava and X-VLM tend to describe images using single nouns without relationship words (97.3\% and 46.0\% in \textit{None} respectively). 
Moreover, all models have a better performance (with a higher score on \textit{Cor} and a lower one on \textit{Exc}) on Attribute Ownership (combination between \textit{two} content words) than Relationship Composition (combination between \textit{three} content words).
From this, it can be observed that VLP models have learned short-distance syntactic combinations and superficial syntactic relationships but struggle to handle complex syntactic relationships.
The performance of comprehending simple semantic features is consistent with the model's performance in the VG-Attribution (superficial) and VG-Relation tasks (complex) in ARO~\cite{yuksekgonul2022and}, where the former achieves a higher score compared to the latter.

In Tab.~\ref{tab: Relationship Composition}, we separately provide the models' performance on each relation category and categorize them into two classes: spatial-based and verb-based relationships~\cite{yuksekgonul2022and}. 
To alleviate the class imbalance problem, we exclude the categories whose sample numbers are more than 25\% of the total samples.
BLIP, CLIP, and X-VLM do not consistently perform badly on sub-categories, particularly regarding verb-based relationships.
In spatial-based relationships, they only exhibit similar performance issues when distinguishing between ``\textit{left}'' and ``\textit{right}''.
Furthermore, in instances where the models demonstrate poor performance, they do not display a preference for selecting \textit{None} classes lacking spatial/verb-based relationships; instead, they tend to choose the \textit{Exc} and \textit{Cor} classes. 
This suggests that the models comprehend the semantics of relationship words using the current syntax but struggle to accurately discern the initiators and recipients of these relationships.

We observe that the performance of VLP models in Tab.~\ref{tab: Relationship Composition} is consistent with their performance on the Semantic Structure~(Tab.~\ref{tab: Semantic_structure}).
When models (BLIP) are sensitive to semantic and syntactic relationships of content words  (with the highest probability choosing ``shuffle non-content words'' and the secondary probability choosing ``Correct'' in Semantic Structure), they become easier to differentiate relationships between nouns (easier to choose \textit{Cor}). 
When models (CLIP) have high sensitivity to syntactic relationships across all words (with the highest probability choosing ``Correct'' in Semantic Structure), they can pose challenges in distinguishing relationships between nouns and verbs and tend to more equitably probable choices of \textit{Cor} and \textit{Exc}.
When models (Flava) only depend on discrete words semantic without syntax (selecting all classes with similar probabilities in Semantic Structure), they tend to opt for the \textit{None} that excludes intricate relationships.
This indicates that syntax is indispensable for the VLP model's acquisition of linguistic knowledge. However, utilizing function words (non-content words) to construct intricate syntactic structures may not effectively enhance the understanding of relationships between words.

\begin{table}[htbp]
\centering
\caption{\textbf{The BLIP2 performance on the Relationship Composition}. BLIP2 demonstrates strong performance on verb-based relationships. However, it exhibits slightly weaker performance on spatial-based ones.
We highlight the sub-categories if the model achieves poor performance (\colorbox{red!10}{\textit{Exc}$>50\%$}). We only show the relationships whose sample numbers are larger than 25.}
\label{tab:blip2-Relationship Composition}
\begin{tabular}{lcccr}
\toprule[1pt]
\multicolumn{1}{l}{Relation} & \multicolumn{1}{l}{\textit{Cor}$\uparrow$} & \multicolumn{1}{l}{\textit{Exc}$\downarrow$} & \multicolumn{1}{l}{\textit{None}$\uparrow$} & \multicolumn{1}{l}{Freq} \\ \midrule[1pt]
\multicolumn{5}{c}{Spatial-based Relationship}                                       \\           
Acc($25<$Freq$<5087$)& 81.9                   & 43.5                   & 98.6                    & 4865                      \\  \hline
above                        & 50.6                   & 27.1                   & 99.3                    & 269                      \\
\rowcolor{red!10} at                           & 88.0                   & 60.0                   & 97.3                    & 75                       \\
behind                       & 80.7                   & 33.6                   & 98.6                    & 574                      \\
below                        & 72.7                   & 39.2                   & 99.0                    & 209                      \\
\rowcolor{red!10} in                           & 82.5                   & 51.7                   & 98.6                    & 708                      \\
\rowcolor{red!10} in front of                  & 91.8                   & 81.6                   & 99.2                    & 588                      \\
inside                       & 87.9                   & 15.5                   & 96.6                    & 58                       \\
\rowcolor{red!10} of                           & 79.6                   & 75.8                   & 98.6                    & 367                      \\
on                           & 85.1                   & 30.6                   & 98.4                    & 1684                     \\
on top of                    & 82.1                   & 14.9                   & 98.5                   & 201                      \\
\rowcolor{red!10} to the left of               & 72.5                   & 73.3                   & 96.4                    & 7741(-)                     \\
\rowcolor{red!10} to the right of              & 73.9                   & 73.3                   & 96.6                    & 7741(-)                     \\
 under                        & 78.0                   & 34.9                   & 98.5                    & 132                      \\ \hline
\multicolumn{5}{c}{Verb-based Relationship}                                                   \\                                         
Acc(Freq$>25$)                 & 88.3                   & 13.8                   & 98.8                    & 1614   \\ \hline
\rowcolor{red!10} covered by                 & 80.6                   & 55.6                   & 97.2                    & 36   \\
covering                     & 78.8                   & 15.2                   & 93.9                    & 33   \\
holding                      & 90.1                   & 5.6                    & 100.0                   & 142  \\
 looking at                   & 77.4                   & 22.6                   & 100.0                   & 31   \\
lying on                     & 86.7                   & 5.0                    & 100.0                   & 60   \\
 riding                       & 86.3                   & 2.0                    & 98.0                    & 51   \\
sitting at                   & 88.5                   & 0.0                    & 96.2                    & 26   \\
 sitting on                   & 89.7                   & 1.7                    & 100.0                   & 175  \\
standing in                  & 84.8                   & 3.4                    & 100.0                   & 59   \\
 standing on                  & 84.6                   & 7.7                    & 100.0                   & 52   \\
wearing                      & 89.4                   & 17.8                   & 98.4                    & 949 \\ \bottomrule[1pt]
\end{tabular}
\end{table}

\subsection{Semantic Level Probing Results}
\label{subsubsec: semantic}
\textbf{
\textit{1)} VLP models cannot distinguish negation logic in multimodal alignment.}
In the Negation Logic task in Tab.~\ref{tab: SNARE}, only the BLIP model achieves an accuracy (79.0\%) above the random level. The performance of CLIP (47.3\%) and X-VLM (48.1\%) is close to the random level. Flava, on the other hand, tends to select sentences with negation words. 
This indicates that it is challenging for the pretraining process to transfer the understanding of negation words from the linguistic knowledge in datasets and language models to VLP models.

\textbf{\textit{\textbf{2)}} VLP models exhibit poor perception of spatial relationships, making it difficult for them to correctly identify simple spatial relationships, especially ``left'' and ``right''.} 
In Tab.~\ref{tab: Relationship Composition}, for ``to the left of'' and ``to the right of'' sub-category in the spatial-based Relationships, all models exhibit similar probabilities of selecting \textit{Cor} or \textit{Exc} ($p<5\%$).  
This suggests that the models are confusing the primary/ secondary objects in the spatial relationships. 
In Fig.~\ref{fig: Spatial_Relation}, we show the model performance on the Multi-spatial Relationship task. To our surprise, all models struggle to differentiate positional relationships. 
In samples whose right relationships match the images are ``to the left of'' and ``to the right of'', the models show similar probabilities distribution to select the four relationship labels and cannot distinguish between ``left'' and ``right''. 
CLIP prefers the relationship ``to the right of'', while Flava prefers ``on''. 
However, these relationships are not the correct options for the given sample.
This indicates that VLP models are mainly incapable of accurately discerning spatial relationships within images or do not understand the positional relationships.
To some extent, it is noteworthy that BLIP and X-VLM showcase the ability to distinguish between ``on'' and ``below'' and otherspatialrelationships in the Relationship Composition task (Tab.~\ref{tab: Relationship Composition}).
During the VLP process, BLIP integrates visual features into the text encoder. In contrast, X-VLM employs a pretraining objective, the Bounding Box Prediction loss, linked to spatial relationships.
This underscores the significance of incorporating spatial-related information during pretraining, which enhances the VLP model's perception of the physical world.

\subsection{Analysis on MLLMs}
\label{subsubsec: MLLMs}
\textbf{
MLLMs exhibit excellent alignment ability and efficiently transfer linguistic knowledge, showcasing their competence in handling intricate syntactic and semantic understanding, including negation comprehension and the composition of triplet relationships.
Nonetheless, MLLMs continue to encounter challenges in accurately learning spatial relationships and exhibit confusion in dealing with the compositional aspects of nouns and attributes within sentences.
}

The Fig.~\ref{fig:blip2_negative}, Fig.~\ref{fig:blip2_attribute_relationshi}, and Tab.~\ref{tab:blip2-Relationship Composition} show the BLIP2 performance on Negation Logic, Attribute Ownership, and Relationship Composition, respectively. 

In the Negation Logic task (Fig.~\ref{fig:blip2_negative}), BLIP2 demonstrates a good understanding of the semantics of negation words when answering questions in the \textit{Wro} class. 
This observation signifies that BLIP2 can comprehend the meaning of negation and engage in reasoning.

However, BLIP2 exhibits similar confusion in the \textit{Cor} class of the Negation Logic task and the \textit{Sep} class of the Attribute Ownership task. (Both of them put nouns behind adjectives in the form of ``\textit{is the $\boldsymbol{X}$ $\boldsymbol{A}$ and is the $\boldsymbol{Y}$ $\boldsymbol{B}$}'' or ``\textit{are the $\boldsymbol{X}$ and the $\boldsymbol{Y}$ $\boldsymbol{A}$ and $\boldsymbol{B}$ respectively}''. It is different from the form of ``\textit{are there the $\boldsymbol{A}$ $\boldsymbol{X}$ and the $\boldsymbol{B}$ $\boldsymbol{Y}$}''. The former emphasizes nouns, while the latter emphasizes attributes.) 
This could be attributed to BLIP2's better alignment of nouns in the multimodal domain but its inability to accurately determine if the attributes belong to the noun. 
In the Attribute Ownership task (Fig.~\ref{fig:blip2_attribute_relationshi}), BLIP2 has a similar performance.
When answering questions in the \textit{Cor} class, BLIP2 accurately determines whether the sentence semantic matches the image. However, even when the nouns and adjectives are mismatched in the \textit{Exc} class questions, BLIP2 still answers ``\textit{yes}'' with a high probability. 
This also indicates that BLIP2 is sensitive to nouns and struggles to judge the semantic combination of attributes and nouns accurately. 
This result highlights the existing limitation that the multimodal alignment in MLLMs is a coarse-grained alignment of entities while overlooking fine-grained alignment.

In Relationship Composition task (Tab.~\ref{tab:blip2-Relationship Composition}), BLIP2 exhibits excellent alignment of nouns (97.0\% and 98.8\% on \textit{None}) and understanding of relationships between entities (75.3\% and 88.3\% on \textit{Cor}), showcasing its visual-language alignment capability and the rich linguistic knowledge obtained from LLMs.
However, it exhibits slightly weaker performance in distinguishing spatial-based relationships (\textit{Exc}) such as ``at'' (60\%), ``of'' (75.8\% ), ``in front of'' (81.6\%), ``to the left of'' (73.3\%), and ``to the right of'' (73.3\%). This suggests that similar to VLP models, MLLMs also encounter challenges in learning precise spatial relationships.

\section{Suggestion}
\label{subsec: suggestion}
LLMs have gained rich linguistic knowledge, physical world knowledge, and impressive reasoning abilities from pretraining on a vast amount of text~\cite{chatgpt, touvron2023llama, scao2022bloom}, thus showing decent performance on a bunch of downstream tasks~\cite{zhong2023chat, Lu2023EAPrompt, Peng2023ChatGPT4MT}. 
However, despite the prevalence of multimodal data in real-life scenarios, acquiring and training high-quality image-text pairs remains challenging. 
Current research on MLLMs addresses this issue by leveraging the knowledge within LLMs to align visual features and textual representations~\cite{yin2023survey}. But MLLMs still have shortcomings in feature representation, comprehension, and reasoning~\cite{fu2023mme}.
To promote the progress of the multimodal community, based on our findings, we offer the following recommendations for future MLLMs research and development:

\textbf{Utilizing LLMs as the language backbone can facilitate the comprehension  of text encompassing intricate semantics, syntax, and logic.}
Comparing the performance of traditional VLP models and the MLLMs on SNARE, the latter can better accomplish multimodal tasks by leveraging the rich linguistic knowledge from LLMs. For example, BLIP2 can understand negation semantics, and its syntactic knowledge can help to construct better complex relationships between words (e.g., the relationship between nouns and adjectives and triplet relationships in the Relationship Composition).

\textbf{Focusing on content words and simplifying the sentence's syntactic structure may be an important approach to constructing high-quality datasets and improving the effect of multimodal alignment.} 
In the Attribute Ownership task, we find that both VLP models and MLLMs easily understand concise syntactic relationships (short-distance combinations between nouns and adjectives). In the Semantic Structure task, VLP models do not exhibit an obvious preference for the sentence without shuffling. In contrast, BLIP, CLIP, and X-VLM prefer sentences that maintain the syntactic relationship between content words. 
This suggests that complex syntactic structures may not be the key factors driving better multimodal alignment performance, and content words are critical elements. 
Hence, employing more content words to establish clearer and simpler syntactic relationships within sentences might contribute to creating higher-quality datasets and enable more effective multimodal alignment.

\textbf{The quality of multimodal modeling largely depends on better fine-grained semantic relationships.} 
Although MLLMs and some VLP models (such as Flava) can accurately align entity features between images and text, they still struggle to differentiate whether the attributes belong to the noun accurately. 
Most VLP models struggle to differentiate verb- or spatial-based relationships between nouns.
Therefore, multimodal models need to learn more fine-grained alignment to structure complex relationships between nouns and attributes accurately. In future work, to improve the fine-grained relationships in the training data, we may explore 1) the refinement of the sentences conditioned on the fine-grained visual information, which can be seen as a kind of knowledge distillation~\cite{Rao2022ParameterEfficientAS, Deng2022ImprovingSM}; and 2) the bidirectional refinement of the paired data~\cite{Ding2021ImprovingNM, Ding2021TheUS}, i.e., refine the text conditioned on the image, and vice versa.

\textbf{Complex visual knowledge mining should be considered in the VLP process.} Both VLP models and MLLMs struggle to accurately determine the spatial relationships of objects in the visual context, especially ``left'' and ``right''. BLIP and X-VLM achieve better results in understanding location information by incorporating visual features into the text encoder and using position-related pretraining objectives. Recently, MLLMs like Kosmos-2~\cite{kosmos-2} have also improved multimodal alignment performance by combining location information into pretraining objectives. Therefore, it is meaningful for future research to explore pretraining objectives that facilitate the learning of fine-grained visual knowledge. Besides, the dynamic learning process, e.g., curriculum learning~\cite{Bengio2009CurriculumL} and progressive learning~\cite{Ding2021ProgressiveMT}, can be employed for the VLP, where the training starts with simple patterns and gradually goes into complex patterns.

\section{Conclusion}
\label{sce:conclusion}
In this paper, we introduce the first comprehensive multimodal alignment probing benchmark -- SNARE for evaluating the impact of linguistic knowledge, e.g., lexicon, syntax, and semantics, for the VLP models. 
We carefully designing four tasks: semantic structure, negation logic, attribute ownership, and relationship composition.
By evaluating the state-of-the-art VLP models, including BLIP, CLIP, Flava, X-VLM, and BLIP2, we show that current VLP models are capable of understanding simple semantics but struggle with complex syntactic relationships and negation logic and lack the modeling of fine-grained information (e.g., spatial relationship and attribute ownership) in visual features.
To enhance the cross-modal alignment modeling, we suggest 1) using LLMs that own rich linguistic knowledge as the language backbone of VLP to improve the understanding and generation of the sentences with difficult semantics and logic, 2) constructing high-quality datasets by closely aligning the visual objectives with the content words in the sentence, and making the syntactic structure simpler, and 3) mining the fine-grained and complex visual knowledge by carefully designing better learning objectives.
We hope that our benchmark and conclusions will facilitate the development of VLP models in the future.

\ifCLASSOPTIONcaptionsoff
  \newpage
\fi

\bibliographystyle{IEEEtran}
\bibliography{reference}

\end{document}